# *PDF/A* standard for long term archiving

**Ramona Vasilescu**
"Tibiscus" University, Timişoara

**ABSTRACT**. *PDF/A* is defined by ISO 19005-1 as a file format based on PDF format. The standard provides a mechanism for representing electronic documents in a way that preserves their visual appearance over time, independent of the tools and systems used for creating or storing the files.

**Introduction**

Archiving formats established by laws and conventions vary from one country to another. Traditional methods of archiving, paper and / or microfilm ensure correct reproduction of documents in the long term. The birth of computers and development of the related technology influenced the setting of standards for archiving using electronic formats for documents.

The first form of electronic long-term archiving was TIFF. This format has the following advantages: it can archive to ensure accurate reproduction and can be easily transmitted over a network of computers, e-mail or through equipment storage.

Disadvantages of this format were revealed in time, as informational techniques have been developed and diversified. Usually TIFF documents are obtained through a scanning process followed by an OCR (Optical Character Recognition).

Additional documents obtained by OCR may be used for contextual searches. OCR processes do not guarantee obtaining the precise scanned text, the result may be accompanied by errors that require further processing by a human user to detect and repair inconsistencies with the original document.

Text correction is essential when the archive is used for text searches. Another weakness of TIFF format is that, being an image, the documents based on this format can be large so it can be difficult to send them at remoteness.

The Adobe Systems Company has launched the PDF file format as a format suitable for situations where electronic documents "migrate" on different operating systems. Over time, PDF has proven to be a file format suitable for archiving for several reasons such as:

- PDF stores structured objects that can be used for fast searches;

- PDF files are compact, the size is smaller than for the TIFF format; so it is easily transmitted to remote;

- type metadata information (such as title, author, subject, keywords) stored in PDF files can be used for automatic classification without human intervention;

- the contents of the PDF page are platform-independent so it is most excellent to use the benefits of reproductive technologies (printers, monitors, etc.) effectively, and the user does not see differences in different content platforms.

Over the past few years, Adobe Systems has released seven versions of the PDF format and each version has been accompanied by: an enrichment and / or update object's structural performance; additional features (such as linearization, to improve loading speed in a web page, implementation of new compression algorithms like *jpeg2000*).

The popularity of this format and frequent changes shaped the idea of defining a new standard for archiving based on PDF format. This new standard was called *PDF/A*.

The initiative to create a standard for electronic archiving based on Adobe PDF format was launched in March 2002 by the Association for Information and Image Management (AIIM), National Printing Equipment Association (NPES) and the Administrative Office of U.S. Courts. This file format, described by means of the ISO 19005-1 standard, provides a mechanism for representing electronic documents in a manner that preserves the visual appearance over time, independent of instruments and systems used for creating, storing and interpreting them.

The standard does not define an archiving strategy or purpose of a system of electronic archiving. ISO 19005-1 may be purchased on site ISO, http://www.iso.org, in paper or electronic file (PDF), written only in English, the documents being under protection of copyright, thus prohibiting the publishing of a free version.

A key element of reproducibility over time is the need for *PDF/A* to be 100% self-contained, i.e. all information needed to display the document (in the same way each time) are embedded in the file. This information includes all information related to the visible text, vector graphics, fonts, color information and more. A *PDF/A* document is not allowed to directly or indirectly use any external information, such as image files, sound files or video files.

## 1. PDF/A versus PDF

Because it is not a standard of the International Organization for Standardization, ISO in short, but it is a company format, PDF format does not guarantee long-term reproducibility and does not meet the "WYSISYG" rule (What you see is what you get).

In order to be accepted, *PDF/A* needed to be based on an existing PDF format described in documentation *Adobe PDF Reference xx* of company Adobe Systems.

For the standard *PDF/A*, ISO TC 171 (TC - Technical Committee) has chosen the documentation Adobe PDF Reference 1.4, which describes the format implemented for Acrobat 5. The standard requires that *PDF/A* complies with all the requirements of PDF documents as amended by ISO 19005. The standard just identifies differences from the reference documentation of PDF. That means PDF 1.4 format documentation should be well understood in order to understand the *PDF/A* standard.

Certain features in PDF 1.4 permits were intentionally excluded from the *PDF/A* (such as some related to sound and video). Some items from *PDF Reference 1.4* are not required or must have certain values. In addition, *PDF/A* requires certain elements to be fully implemented, such as embedded fonts. In short, *PDF/A* is based on the *PDF Reference 1.4* documentation, with features that may be mandatory, recommended, prohibited, or restricted.

*PDF/A* is a standard that has two parties for approval and implementation. Until now, only the first *PDF/A*-1 has been approved. The latest part two, *PDF/A*-2, will cover new features added in versions 1.5, 1.6 and 1.7 of the *PDF Reference* documentation. *PDF/A*-2 should ensure compatibility with previous versions, i.e. all valid *PDF/A*-1 must comply with *PDF/A*-2, but the files that do not comply with *PDF/A*-2 need not comply with the *PDF/A*-1.

*PDF/A*-1 is divided on two levels of compliance: *PDF/A*-1a and *PDF/A*-1b. *PDF/A*-1a (referred as Level A compliance) designates full compliance with Part I of the current standard *PDF/A* ISO 19005-1. Minimal level of compliance is defined by *PDF/A*-1b (referred to as Level B compliance). *PDF/A*-1b requirements were intended to ensure that the visual appearance of the file is reproducible in the long run.

Among the differences between standards *PDF/A*-1a and *PDF/A*-1b, notable are those relating to the extraction of text:

- *PDF/A*-1a ensures preservation of the document logical structure and texts in natural reading order. Text extraction is important when the document must be viewed on the screen of a mobile device or through other devices for which the text should be reorganized according to the data limitations of screen size (through the so-called re-flow processing). This facility is recognized as the tagged PDF;

- *PDF/A*-1b ensures that text and additional content can be displayed correctly, but does not guarantee that text will be extracted in natural reading order, or that it will be readable or logical.

Differences between *PDF/A*-1a and *PDF/A*-1b have no impact on scanned documents, if the files have not provided updates for OCR processing purposes (e.g. to obtain text search).

Documentation of standard *PDF/A* is short, but very technical, and the associated documentation is vast. Therefore, the standard can be understood on the stark by experts who have fundamental knowledge on language description pages such as PostScript and PDF.

The PDF reference documentation contains more than 1000 pages. Further references relating to the structure of fonts, XML specification, compression formats and so forth may be studied. For a company producing software, a good approach is to hire an expert who can determine the need for *PDF/A*, how to implement the *PDF/A* archiving strategy and detail the steps needed to fully accomplish the objective of archiving.

Other issues to be considered when implementing an archive that complies with the standard *PDF/A* are related to organizational standards and procedures, sources of consistent data, font reliability, quality management and other specific needs. Migrating from the current archive (either paper or TIFF) to *PDF/A* archiving is not an easy task to do and must be planned carefully.

The market is not flooded with *PDF/A* products. Understanding the technology behind the *PDF/A* requires extensive knowledge, and users have the highest expectations of the software according to the standard. The first instruments have reached the market in 2006. Most are designed in order to create, convert and check whether the existing files comply with the standard (see Table 1).

**Table 1. Software for PDF files conform *PDF/A* the standard**

| Producer | Product | Use |
| --- | --- | --- |
| LuraTech | LuraDocument PDF Compressor | creating files |
| LuraTech | LuraDocument PDF/A Printer | creating files |
| LuraTech | LuraDocument PDF Validator | checking files |
| callas software | pdfaPilot Validator CLI | checking files |
| callas software | pdfaPilot Converter CLI | converting files |
| callas software | pdfPilot Plug-in | checking and converting files |
| PDF Tools AG | 3-Heights PDF Producer | creating files |
| PDF Tools AG | 3-Heights Image to PDF Converter | creating files |
| Tracker Software Products | PDF Xchange | creating files |
| May Software | eDocPrintPro PDF/A | creating files |

*PDF/A* is not seen as just a trend. The devices for archiving have developed during the past few years; PDF is already established as a type of archiving. The *PDF/A* standard will help ensure long-term preservation of electronic documents

Description of the *PDF/A* standard is short, presented in a form that requires knowledge and references to other principles such as:

- Adobe documentation on the format of the structure of 1.4 PDF;

- documentation on the formats for calendar data and time;

- specification Extensible Markup Language (XML) and RDF / XML (RDF - Resource Description Framework);

- specification for file format of color profiles.

The description of a property of a PDF document from the catalog must correspond with the description of the metadata associated with the property. Converting a PDF file to the PDFA format will requires the ICC profiles (International Color Consortium). The ICC profiles are used to obtain correct colors for another color space. These profiles can be obtained from the website of Adobe, which provides two packages:

- for end users (the package "ICC profile download for End Users");

- for users who intend to redistribute ICC profiles through hardware and software ("ICC profile download for Bundling")

Both packages can be downloaded free of charge, the difference between these packages lies in how to use licensing. These ICC profiles are necessary for building dictionaries that describe color characteristics through the *Output Intents* dictionary.

**Conclusions**

PDF is used worldwide for private and public documents and is accepted as a format for archiving in countless markets. The Adobe Systems Company will continue to improve opportunities and technologies related to PDF (e.g. 3D and XFA for dynamic forms). However, the *PDF/A* standard will remain constant and usually will not change, the purpose of this standard being to ensure long-term reproducibility. Most likely, the standard *PDF/A* will have an impact on future development of the PDF format and the Adobe company should consider the format of *PDF/A* and not "lose" certain elements and features of future versions of PDF format.

Market software for *PDF/A* is not a rich one, but is in the process of enlargement. Given the knowledge gained by archiving in various formats, over time it was found that different formats were not functional in the present, and the applications for processing have changed so much that the archives have become unusable.

Thus, the predominant trend is to use a file format which guarantees that it will change over time, even if it is for 5 or 10 years. We can consider this format to be PDF. Currently, the interest for standard *PDF/A* and its use in Romania is not developed, even more limiting the announcement that some virtual printers can obtain a PDF file that complies with the *PDF/A* standard.